\documentclass[twocolumn]{aastex701}

\usepackage{comment}
\begin{document}

\title{Little Red Dot progenitors from Compact Starbursts: A Natural Path to Early AGN Formation}

\author[orcid=0009-0006-9227-1343,sname='Liempi', gname='Mat\'ias']{Mat\'ias Liempi}
\affiliation{Dipartimento di Fisica, Sapienza Universit\`a di Roma, Piazzale Aldo Moro 5, 00185 Rome, Italy}
\email{gonzalez.liempi@uniroma1.it}  

\author[orcid=0000-0003-2480-0988]{Muhammad A. Latif} 
\affiliation{Physics Department, College of Science, United Arab Emirates University, PO Box 15551,Al-Ain, United Arab Emirates}
\email[show]{latifne@gmail.com}

\author[orcid=0000-0002-9642-120X]{Dominik R.G. Schleicher}
\affiliation{Dipartimento di Fisica, Sapienza Universit\`a di Roma, Piazzale Aldo Moro 5, 00185 Rome, Italy}
\email{dominik.schleicher@uniroma1.it}

\begin{abstract}
The recent discovery of Little Red Dots (LRDs) by the James Webb Space Telescope has challenged traditional models of early galaxy and black hole co-evolution. The nature of these highly compact objects remains heavily debated, with explanations divided between dust-reddened active galactic nuclei (AGN) and extremely dense stellar populations. We perform high-resolution cosmological simulations  to model the formation of LRD precursors. Motivated by recent high-redshift observations and theoretical results, we specifically explore environments characterized by high star formation efficiencies (30\% and 100\%) and confined feedback. Our simulations naturally produce highly compact galaxies with stellar masses of $10^7-6 \times 10^8 $\,M$_\odot$, with most of the mass concentrated within $200-300$ pc. We find that, in these dense environments, gas inflows, gravitational torques, and stellar dynamical friction operate on highly efficient timescales. Over a 10 Myr timescale, gas inflows can accumulate $\rm \sim 10^7 M_\odot$ at the galactic center, while gravitational torques and dynamical friction can contribute an additional $10^5-10^9$\,M$_\odot$ and $10^3-10^4$\, M$_\odot$  through the inward migration of massive stars. Assuming a conservative 10\% efficiency to account for feedback, this rapid mass accumulation can lead to the formation of a $\sim 10^6$\,M$_\odot$ central black hole, naturally giving rise to an AGN in these dense systems. Therefore, stellar and AGN interpretations of LRDs may not be mutually exclusive; rather, dense stellar systems are likely precursors to AGN.
\end{abstract}


\section{Introduction}

The emergence of Little Red Dots (LRDs) in {\it James Webb} Space Telescope (JWST) surveys, particularly the JWST Advanced Deep Extragalactic Survey (JADES) \citep{Eisenstein2025, Eistenstein2026}, has challenged existing paradigms of early galaxy and black hole co-evolution. These objects are primarily identified by their "V-shaped" spectral energy distributions (SEDs)—characterized by a blue rest-frame ultraviolet (UV) continuum and a steeply rising red rest-frame optical continuum—and their extremely compact, often point-like morphologies \citep[e.g.,][]{KOCEVSKI2023,MATTHEE2023, GREENE2024, AKINS2024}. 

However, the nature of LRDs remains a subject of heavy debate, centered on whether their luminosity is driven by dust-reddened active galactic nuclei (AGN) or extremely dense stellar populations.  Under a purely stellar interpretation, LRDs exhibit stellar masses ranging between $10^8$ and $10^{11}$\,M$_{\odot}$ \citep{GREENE2024} in an effective radius of $3 - 300$\,pc \citep{MATTHEE2023,AKINS2024}. If such stellar masses are contained in such compact volumes, LRDs may have median core densities about $10^4$\,M$_\odot$\,pc$^{-3}$ with peak values reaching $\sim 10^8$\,M$_{\odot}$\,pc$^{-3}$ \citep{GUIA2024}. Such scenarios would imply that these objects have rather short collision timescales but long relaxation times, potentially leading to the formation of very massive central objects through collision-based channels \citep{Escala2025, Pacucci2025, Liempi2026}. Indeed, LRDs exhibit AGN indicators such as broad Balmer emission lines (e.g., $H_{\alpha}$,$H_{\beta}$) with FWHM $>1000$~km\,s$^{-1}$ that are associated with the broad line region \citep{TAYLOR2025A}, implying the existence of a supermassive black hole (SMBH) accreting at the center.

The limitations of both the purely stellar and standard AGN models have prompted the exploration of more exotic configurations. Recent studies suggest that LRDs may represent SMBHs enshrouded by extremely dense, compact gas envelopes. This model explored by \citet{RUSAKOV2026} and \citet{SNEPPEN2026} demonstrates that the unique "V-shaped" spectra of LRDs can be reproduced by a central engine buried within a dense turbulent gaseous medium. This environment obscures the accretion disk while potentially giving rise to the broad emission lines observed in recent surveys  \citep[e.g.,][]{PACUCCI2024}.

Complementary to the obscured AGN hypothesis is the possibility that LRDs are the observational signatures of Supermassive Stars (SMS). In this scenario, the "red" component is not necessarily dust-reddened AGN light but the photospheric emission of a massive, non-equilibrium star. Recent modeling by \citet{ZWICK2025} and \citet{NANDAL2026} suggests that the SEDs of LRDs are consistent with the properties of these short-lived stars. 

An additional hint regarding their nature may come from the fact that two LRDs have been observed recently within one $z\sim7$ galaxy at about $70$~pc separation \citep{Yanagisawal2026}. This may suggest that mergers could be involved in their formation, as for example outlined by \citet{Dekel2025}; or they might just occasionally go through merger events, or perhaps there can be astrophysical situations in high-redshift galaxies where the red components provide Lyman Werner flux comparable to those required for direct collapse BHs \citep{Bag26}.

Even if the exact nature of LRDs is not fully clear, their possible explanations point either towards an existing SMBH or an SMBH about to be formed. Observations at $z \sim 7-9$ reveal stellar masses of $10^{10}$\,M$_{\odot}$ concentrated within sub-kiloparsec scales \citep{BAGGEN2023, FURTAK2023}. At these extreme densities, the dynamical timescales for stellar relaxation and collision become shorter than the lifetimes of the most massive stars, triggering runaway merger events. Current theoretical frameworks suggest that such environments facilitate the birth of 'heavy' black hole (BH) seeds via the collapse of supermassive stars or direct gas-cloud collapse \citep{Begelman2010,  Schleicher2013, Haemmerle2020}. Numerical simulations have strongly assessed the possible formation pathways of SMBHs, including direct collapse \citep{Bromm2003, Latif2013a, Latif2013b, Latif2022, Mayer2024}, runaway stellar collisions \citep{Portegies2004, Reinoso2018, Escala2021, Vergara2023, VERGARA2024, Vergara2025, Vergara2026, LIEMPI2025, Rantala2025, Rantala2026}, gas inflows into black hole clusters \citep{Lupi2014, Kroupa2020, Gaete2024} as well as mixed scenarios considering the interplay between accretion and collisions \citep{Boekholt2018, Tagawa2020, Chon2020, Chon2025,  Schleicher2023, Reinoso2023, Reinoso2025,Saavedra2024,  Solar2025}. 

For the high-redshift galaxy evolution, JWST in general has found a rather shallow evolution of the volume densities \citep{Chworowsky2024}, a phenomenon that has been interpreted as a result of enhanced star formation efficiencies (SFEs) in the high density gas at high redshift \citep{Somerville2025}. Such an enhancement would be physically expected and is consistent with cloud-scale simulations \citep[e.g.,][]{Kim2018, Lancaster2021, Menon2025}. Similarly, it was found that the efficiency of radiative and mechanical feedback decreases with the density of the environment \citep{Haid2018}. At high redshift, the feedback in galaxies within a relevant mass range is expected to be confined, allowing for strong starbursts without significant amounts of feedback \citep{Dekel2023, Dekel2025, Yajima2025}.

While large cosmological simulations successfully reproduce the macroscopic properties of  LRDs with masses $\rm \geq 10^{8}~M_{\odot}$ at $z \sim 5-8$ \citep{LaChance2025a, LaChance2025b}, they can not follow the assembly of their progenitors at earlier times. Resolving the localized mechanisms that seed these extreme environments at earlier epochs requires targeted models. Here, we present high-resolution zoom-in cosmological simulations exploring the regime of high SFE and confined feedback. Studies using direct N-body models explored the detailed internal dynamics of isolated dense star clusters demonstrating how runaway collisions, tidal disruption events (TDEs), and binary interactions efficiently grow central massive black holes \citep{SAKURAI2019, RIZZUTO2023, ARCASEDDA2024A, Rantala2024}.  Rather than resolving star-by-star interactions within pre-existing systems, this study focuses on the formation and evolution of compact systems in  cosmological environments at earlier times ($z>10$).

Our goal is to model the formation of a galaxy with a stellar mass of $10^7-10^8$~M$_\odot$, roughly corresponding to the low-mass end of LRD progenitors. We show here that this quite naturally leads to the formation of quite compact galaxies, where most of the gas and stellar mass is located within $\sim200$~pc. We further demonstrate that such a configuration implies short timescales for dynamical friction and gravitational torques, implying the likely formation of a central massive object and an AGN within less than a Hubble time. The configurations found here could thus be considered as natural progenitors of LRDs, depending on the scenario under consideration.

In section~\ref{sec:method}, we present our methodology for the cosmological simulations. The results of the simulations are presented in section~\ref{sec:results}, including an analysis of the migration timescales to the center of the galaxies. A summary with our main conclusions is provided in section~\ref{sec:summary}.

\section{Methodology} \label{sec:method}

The numerical cosmological hydrodynamics simulations presented here were conducted with the open source code \textsc{ENZO} \citep{enzo} using a 3rd order piece-wise parabolic solver for hydrodynamics, a particle-mesh method to solve $N$-body dark matter (DM) dynamics and a multigrid Poisson solver for to calculate self-gravity.  We use cosmological parameters from \textit{Planck-2} \citep{planck2} $\Omega_{\mathrm{M}}=$ 0.3089, $\Omega_{\Lambda}=$ 0.691, $\Omega_{\mathrm{b}} = $ 0.0486, $h =$ 0.677, and $n =$ 0.967.
  
Our simulations started at $z=200$ from cosmological initial conditions generated from MUSIC \citep{Hahn11} in a periodic box of $37.3$~Mpc with a top grid resolution of $512^3$. We first run simulations with a uniform grid resolution of $\rm 512^3$ down to redshift $6$ and selected the most massive halo forming in the given volume using the Rockstar halo finder \citep{rockstar}, identifying DM particles belonging to the Lagrange volume of the halo and tracing them back to the initial redshift. We then added one refinement level in a region twice the Lagrange radius of the halo and rerun the simulation. This resulted in a DM resolution of  $\rm  \sim 10^6 ~ M_{\odot}/h$. We smoothed DM particles at refinement level 8, corresponding to physical scales of ~16 pc, to mitigate spurious heating by dark matter particles.  This smoothing effectively suppresses the two-body scattering that drives artificial collisional heating between DM and baryons \citep{Wise2008,enzo,Latif2022,L26}. Our selected halo has a mass of $\rm 3 \times 10^{10}~ $M$_{\odot}$ at $z=11$ ($3.5\times10^{11}$~M$_\odot$ at $z=6$).

During the course of the simulations, we further employed 10 dynamical levels of refinement to resolve the gravitational collapse down to  scales of about $4$~pc. Our refinement criteria are based on the baryonic overdensity, DM mass resolution and  the Jeans refinement of 4 cells. For  details we refer to \citet{latif20b} and \citet{latif22b}. To solve the thermal and chemical evolution of the gas along with the hydrodynamics, we employ the non-equilibrium chemistry network, which solves the rate equations of six primordial species $\rm H, ~H^+, ~He,~ He^+, ~He^{++},and ~e^-$ \citep{anet97}. Our chemical model includes cooling due to collisional excitation and  collisional ionization, radiative recombination Bremsstrahlung radiation and the Compton heating/cooling. We further assume here that the progenitors of the simulated halo are enriched by metals from the first generations of stars forming in minihalos which here in our simulation are unresolved. Therefore, we explored here the implications of two fixed metallicities of $0.01$ and $0.1$~${\rm Z }/{\rm Z}_{\odot}$, assuming an overall efficient metal enrichment related to the high SFE in the environment under consideration. Our chemical model includes metal line cooling  (C, O, N, Si, etc) following \citet{japp07} valid for temperature of $\rm 100-10^4~K$  and uses the metallicity dependent tabulated cooling function of \citet{Sutherland93} for temperatures $\geq 10^4$ K.

Our recipe for star formation is based on \citet{CO92} with no-delayed star formation and creates star particles in  grid cells when meeting the following criteria; 1) the cell is at the maximum refinement level; 2) the gas density in the cell is higher than $\rm 10^4 \, cm^{-3}$; 3) the flow is converging (divergence is negative); 4) the cooling time is shorter than the dynamical time , 5) the star particle mass that would form is greater than $100$ solar masses. In comparison with traditional approaches, we do not impose stochastic star formation and also relax the Jeans instability condition, which allows to accumulate more mass than the Jeans mass as not all cloud scales are being well resolved. For further details, see \citet{kim11}. Considering the very high SFEs implied by JWST observations of high-redshift galaxies \citep{Somerville2025}, we adopted SFEs of 30 \% and also studied a case with $100\%$. While in typical environments, self-regulation via stellar feedback decouples the global SFE from local sub-grid parameters, the high-redshift Universe presents a distinct regime. Due to cosmic evolution, DM halos at $z > 8$ have higher average virial densities, resulting in more compact systems with deeper gravitational potentials that are inherently more robust against feedback. Indeed, theoretical analyses by \citet{Dekel2023, Dekel2025} demonstrate that in halos of $\sim 10^{10.5}\,{\rm M}_{\odot}$ at these redshifts—matching the conditions in our simulated halos—supernova feedback is highly confined. \citet{Yajima2025} come to very similar conclusions based on criteria designed to assess the efficiency of feedback on these systems. \citet{Haid2018} pursued a more general exploration regarding the efficiency of feedback and how it depends on density that supports the results from the other studies.  The gas surface densities obtained in our zoom-in regions place these systems directly in the regime where SFEs of order unity are expected, consistent with the cluster-scale analysis of \cite{Somerville2025}. Therefore, we consider feedback to be confined in our primary runs. We explicitly state this as an assumption designed to explore this novel, high-density regime, and we intentionally vary the SFE and include a comparative case with supernova feedback (the "FB" run) to explore the associated uncertainties. For the "FB" run we turn on thermal and chemical feedback from  star particles of type Ia supernova. The mass loss and luminosity of supernovae are determined from the fits \footnote{\url{https://www.physics.unlv.edu/~kn/SNIa_2/}} of \citet{Nag04}

\section{Results} \label{sec:results}

We present the global properties of the compact galaxies formed in our simulations in section~\ref{global}. In section~\ref{timescale}, we pursue a timescale analysis applied to the simulation data showing the expected transport of gas and stellar mass into the center of the galaxy due to  gravitational torques and dynamical friction, suggesting the likely formation of a central massive object and the development of an AGN.

\begin{deluxetable*}{lcccc}
\tabletypesize{\footnotesize}
\tablewidth{0pt}
\tablecaption{Summary of the simulation parameters explored here. \label{tab:sims}}
\tablehead{
\colhead{Simulation} & \colhead{Star formation efficiency} & \colhead{Feedback} & \colhead{Metallicity} & \colhead{Redshift} \\
\colhead{} & \colhead{(\%)} & \colhead{} & \colhead{(${\rm Z}_\odot$)} & \colhead{($z$)}
}
\startdata
SFE = 30\% & 30 & No & 0.1 & 13.0 \\
SFE = 30\% LowZ & 30 & No & 0.01 & 14.9 \\
SFE = 30\% FB & 30 & Supernova & 0.1 & 12.6 \\
SFE = 100\% & 100 & No & 0.1 & 12.5 \\
\enddata
\end{deluxetable*}

\begin{figure*}
\begin{center}
\includegraphics[scale=0.45]{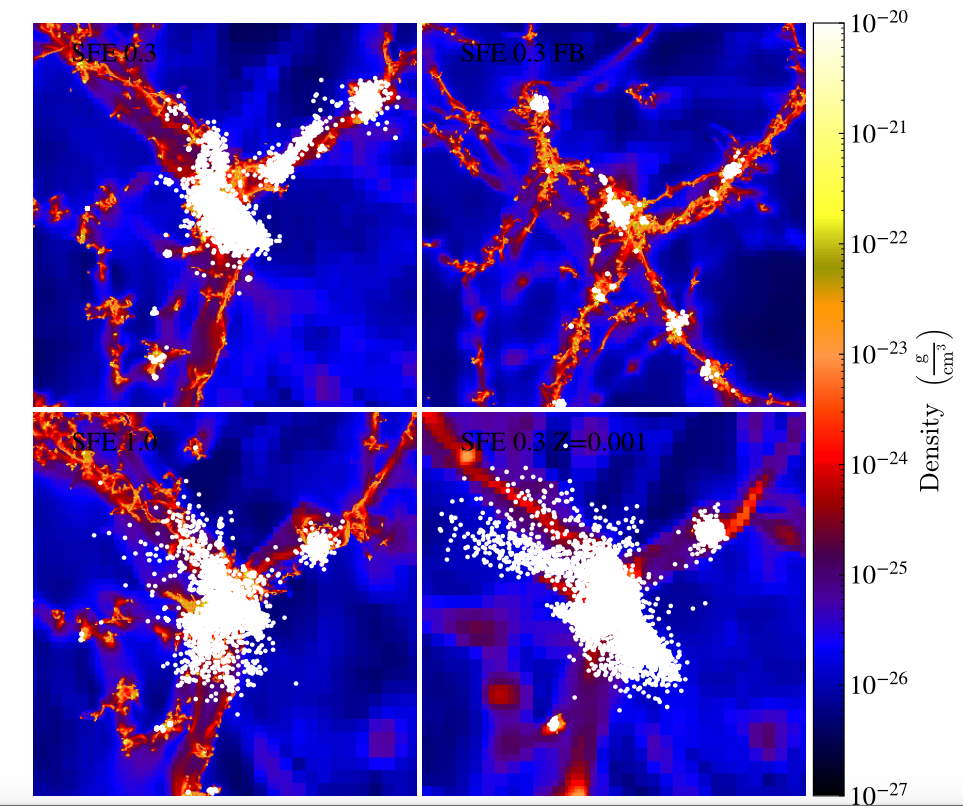} 
\end{center}
\caption{Density projections showing the average density along the line of sight for all four runs for the central 10 kpc region. White dots represent the Pop II star particles.}
\label{fig:spec}
\end{figure*}

\begin{figure*}
\begin{center}
\includegraphics[scale=0.6]{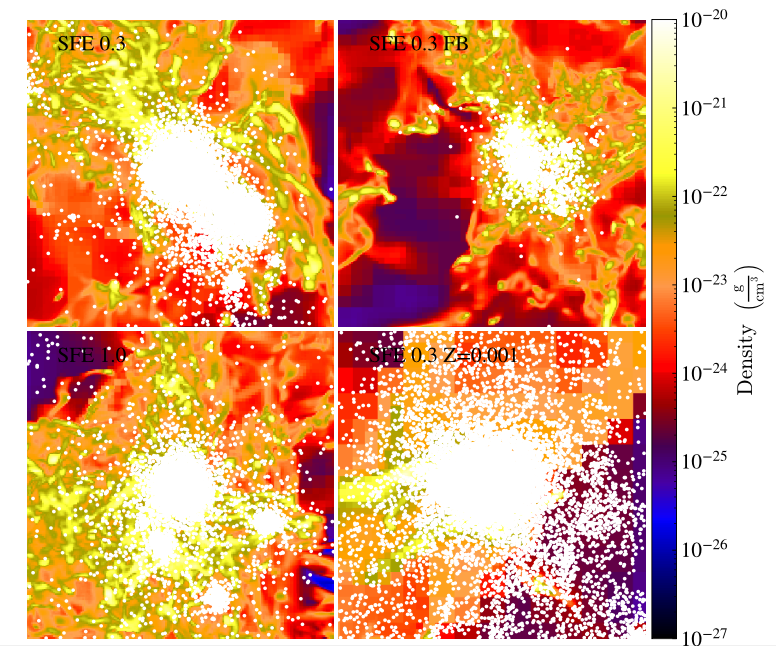} 
\end{center}
\caption{Same as figure 1, showing the gas and stars distribution on 1 kpc.}
\label{fig:1kpc}
\end{figure*}

\begin{figure}
    \centering    \includegraphics{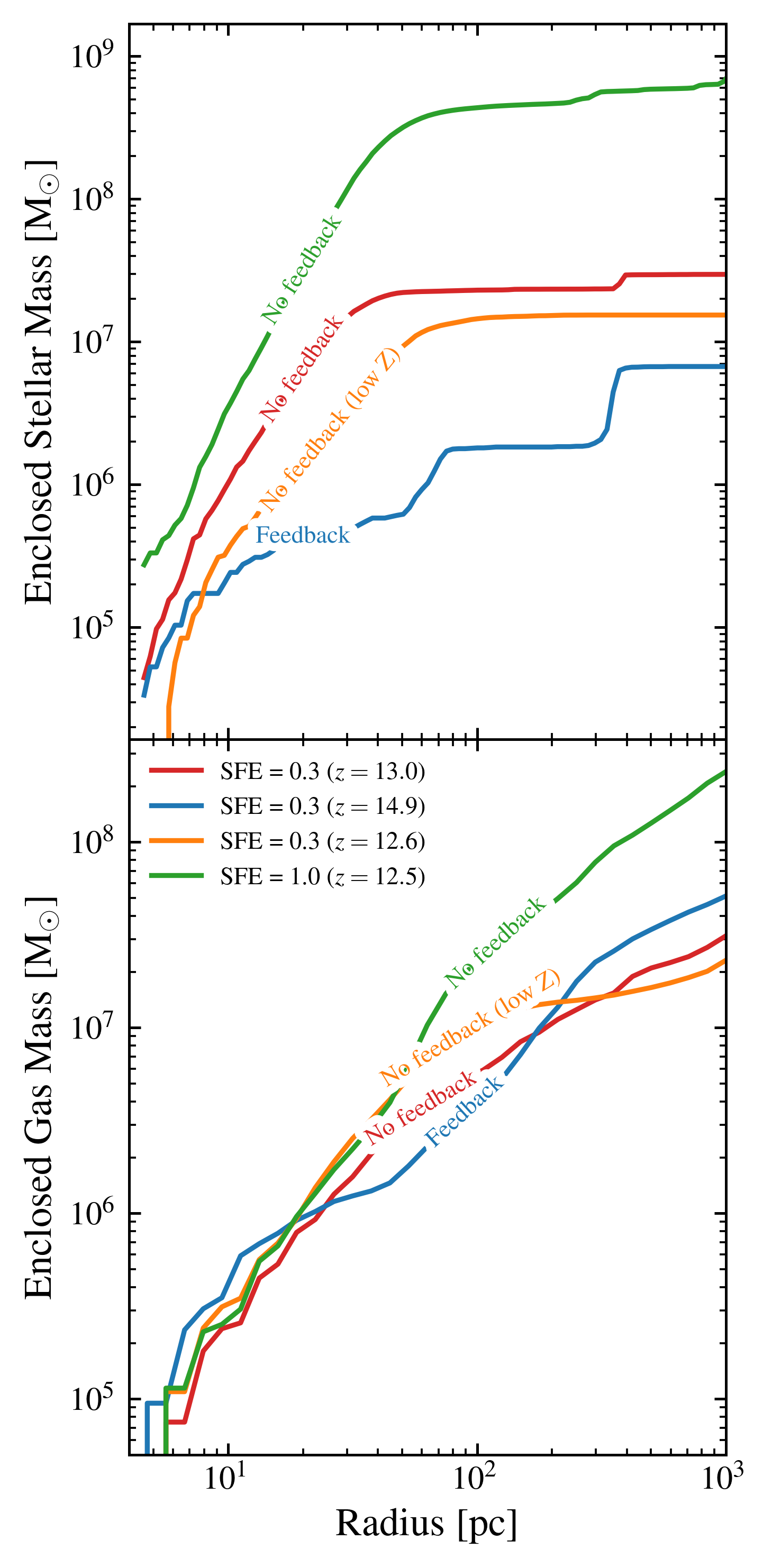}
\caption{
Radial profiles of the enclosed stellar mass (top panel) and enclosed gas mass (bottom panel). The curves compare four simulated configurations: A model with a SFE of $30\%$, no feedback, and $0.1$\,Z/Z$_\odot$ (red); a model with supernova feedback, SFE of $30\%$, and  $0.1$\,Z/Z$_\odot$  (blue); a low-metallicity model with $0.01$\,Z/Z$_\odot$, and no feedback (orange); and an extreme SFE model of $100\%$, no feedback, and metallicity $0.1$\,Z/Z$_\odot$ (green).}  \label{fig:MassProfiles}
\end{figure}

\begin{figure*}
    \centering
    \includegraphics{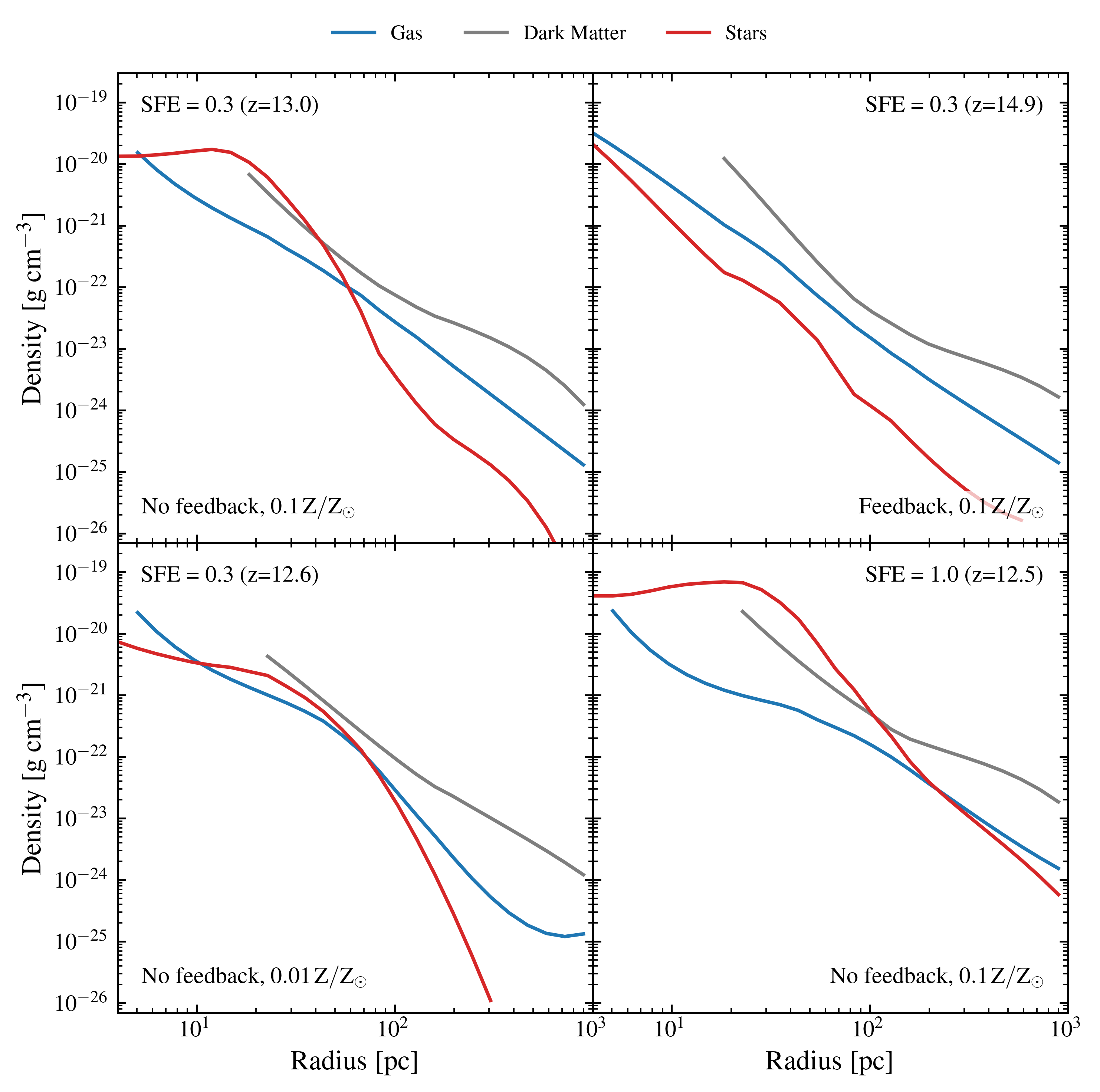}
    \caption{Radial profiles of DM density, gas density and stellar density for the different Simulations. Top left: SFE of $30\%$, no feedback, metallicity $0.1$~Z/Z$_\odot$. Bottom left: SFE of $30\%$, no feedback, metallicity $0.01$~Z/Z$_\odot$. Top right: SFE of $30\%$, supernova feedback, metallicity $0.1$~Z/Z$_\odot$. Bottom right: SFE of $100\%$, no feedback, metallicity $0.1$~Z/Z$_\odot$.    }
    \label{fig:radialprofiles}
\end{figure*}

\subsection{Global results}\label{global}

We have pursued a set of cosmological simulations to explore the formation of a high-redshift protogalaxies with stellar masses of $10^7-10^8$~M$_\odot$. For our reference run, we have considered parameters motivated from JWST observations, in particular with an increased SFE of $30\%$ and a metallicity of $0.1$~Z$_\odot$ reflecting the expected rapid metal enrichment in such an environment. High SFEs are physically expected in dense environments \citep{Somerville2025} and fully consistent with the results from cloud-scale simulations \citep{Kim2018,Lancaster2021, Menon2025}. Given the high SFEs, we implicitly assumed that feedback will be less relevant \citep[see also][]{Haid2018, Dekel2023, Dekel2025, Yajima2025} and neglected it in the majority of the simulations. We varied the simulation parameters, including cases with a SFE of 100\%, a lower metallicity of 0.01 Z$_\odot$, and a comparison run with supernova feedback. A summary of the configurations we explored is given in Table~\ref{tab:sims}. The simulations have been evolved until further computation became computationally prohibitive, leading to final redshifts between $12.5$ and $14.9$. We thus note that they did not reach exactly the same evolutionary stages; nonetheless, it is striking that they still share several common properties.

We show the average density along the line of sight at  the final stages of the simulations for the $10$~kpc region in Fig.~\ref{fig:spec}. Three streams of cold dense gas from large scales (beyond the virial radius of the host halo) are feeding galaxy. Most of the stars form in the central halo, see also Fig.~\ref{fig:radialprofiles}, but some form in the dense filaments and gets merged in the center over time. As discussed below most of the stellar mass lies in central a few hundred pc. A more zoomed-in version of the gas and stellar mass distribution is given in Fig.~\ref{fig:1kpc}, showing them on a $1$~kpc scale. A central concentration of the stellar distribution as well as the presence of sub-structures is clearly recognizable in the different simulations.

The radial profiles of the enclosed stellar and gas masses for all runs are shown in Fig.~\ref{fig:MassProfiles}. The enclosed stellar mass varies within the simulation but they have a common feature that most of the enclosed stellar mass is between $200-300$~pc. 
As expected, we obtain the largest stellar mass in the simulation with $100\%$ SFE with $\sim6\times10^8$~M$_\odot$ on $300$~pc scales. The two runs with SFEs of $30\%$ produce stellar masses of $2-3\times10^7$~M$_\odot$ on the same scale, while in the presence of supernova feedback the obtained stellar mass is $\sim7\times 10^6$~M$_\odot$. In the runs without feedback, the enclosed stellar mass increases rather steeply, almost as radius $R^{3/2}$, while in the run with feedback the increase is initially more shallow and becomes rather steep in the few $100$~pc range. This is due to the supernova feedback from stars which on one hand heats gas but simultaneously eject metals resulting in metallicities of up to a few times solar. Such high metallicity increases the gas cooling and results in stellar mass a factor of 2-3 smaller than the run without feedback. The numbers of course need to be compared with a grain of salt, since they correspond to slightly different cosmological redshifts varying from 14.9-12.5. All runs show a significant stellar mass of $3\times10^5-2\times10^6$~M$_\odot$ on $10$~pc scales. Overall, we find that a compact galaxy forms in all four runs irrespective of SFE and feedback which may later form an LRD.

We further assess the amount of gas in the galaxy for all four runs. The enclosed gas mass on scales of $100$~pc corresponds to $\sim4\times10^6-3\times10^7$~M$_\odot$, providing gas surface densities consistent with the high SFEs assumed here \citep{Somerville2025}. As expected, the lowest enclosed gas mass is obtained in the run with supernova feedback on the $100$~pc scale and the largest enclosed gas mass is found in the run  with $100\%$, which has also been evolved to the lowest redshift of $12.5$. The simulation with $30\%$ SFE and metallicity $0.01$~Z$_\odot$ has reached a similar redshift of $12.6$, but showing an intermediate enclosed gas mass of order $8\times10^6$~M$_\odot$, suggesting that the cooling efficiency could be a further relevant factor  regulating the enclosed gas mass. We note that towards smaller scales, the different simulations show a rather similar enclosed gas mass of $\sim10^6$~M$_\odot$ around $20$~pc. The gas mass scales approximately as the radius, thus approximately corresponding to an isothermal sphere. A summary of the enclosed gas and stellar masses of the different simulations within $10$~pc and $100$~pc is given in Table~\ref{tab:enclosed_masses}.

\begin{deluxetable}{lrrrrr}
\tabletypesize{\footnotesize}
\tablewidth{0pt}
\tablecaption{Enclosed gas and stellar masses within $10$~pc and $100$~pc for the different simulations. \label{tab:enclosed_masses}}
\tablehead{
\colhead{Simulation} &\colhead{redshift} &\colhead{Radius} & \colhead{$M_{\rm stellar}(<R)$} & \colhead{$M_{\rm gas}(<R)$} & \colhead{$M_{\rm tot}(<R)$} \\
\colhead{} & 
\colhead{} &
\colhead{[pc]} & \colhead{[M$_\odot$]} & \colhead{[M$_\odot$]} & \colhead{[M$_\odot$]}
}
\startdata
SFE = 30\% & $13.0$ & 10 & $1.02 \times 10^{6}$ & $2.45 \times 10^{5}$ & $1.27 \times 10^{6}$ \\
SFE = 30\% LowZ & $14.9$  & 10 & $3.54 \times 10^{5}$ & $3.25 \times 10^{5}$ & $6.79 \times 10^{5}$ \\
SFE = 30\% FB & $12.6$  & 10 & $2.27 \times 10^{5}$ & $4.17 \times 10^{5}$ & $6.45 \times 10^{5}$ \\
SFE = 100\% & $12.5$ & 10 & $3.48 \times 10^{6}$ & $2.69 \times 10^{5}$ & $3.75 \times 10^{6}$ \\ \hline
SFE = 30\% & $13.0$ & 100 & $2.30 \times 10^{7}$ & $5.69 \times 10^{6}$ & $2.87 \times 10^{7}$ \\
SFE = 30\% LowZ & $14.9$  & 100 & $1.45 \times 10^{7}$ & $1.03 \times 10^{7}$ & $2.48 \times 10^{7}$ \\
SFE = 30\%  FB & $12.6$ & 100 & $1.81 \times 10^{6}$ & $3.99 \times 10^{6}$ & $5.80 \times 10^{6}$ \\
SFE = 100\% & $12.5$ & 100 & $4.36 \times 10^{8}$ & $2.27 \times 10^{7}$ & $4.59 \times 10^{8}$ \\
\enddata
\end{deluxetable}

To evaluate the global properties of the simulations, we show the density profiles of their gas, stars and DM in Fig.~\ref{fig:radialprofiles}. The DM density is found to be the dominant component up to scales of about $100$~pc and scales approximately as the inverse of the radius squared. The gas density, as previously discussed resembles approximately an isothermal sphere. 
Some deviations are seen in the lower-metallicity run, where reduced cooling may limit collapse efficiency, and in the 100\% SFE run, where rapid gas conversion into stars leads to very high central stellar densities. In the 100\% SFE run, the stellar component dominates within 100 pc, reaching a peak density of 1000 M$_\odot$ pc$^{-3}$ at 20 pc, followed by a mild decline toward smaller radii. A similar trend is seen in the run with 30\% SFE and 0.1 Z$_\odot$, where the stellar density exceeds the dark matter density within 30 pc, though the peak is lower by a factor of $5$, reflecting differences in SFE and final redshift. For the simulation with lower metallicity of $0.01$~Z$_\odot$ and SFE of $30\%$, the stellar density essentially remains comparable to the gas density and the system is somewhat less evolved due to the lower cooling efficiency, despite a final redshift of $12.6$ being reached. In the run with supernova feedback, on the other hand, the peak stellar density is comparable to other cases, it sharply increases towards the center at scales of below 100 pc but remains lower than the gas density at all scales. Across all simulations, the stellar density in the central 10 pc ranges from $40.5$ to $675$ M$_\odot$ pc$^{-3}$.

\begin{figure}
    \centering
    \includegraphics{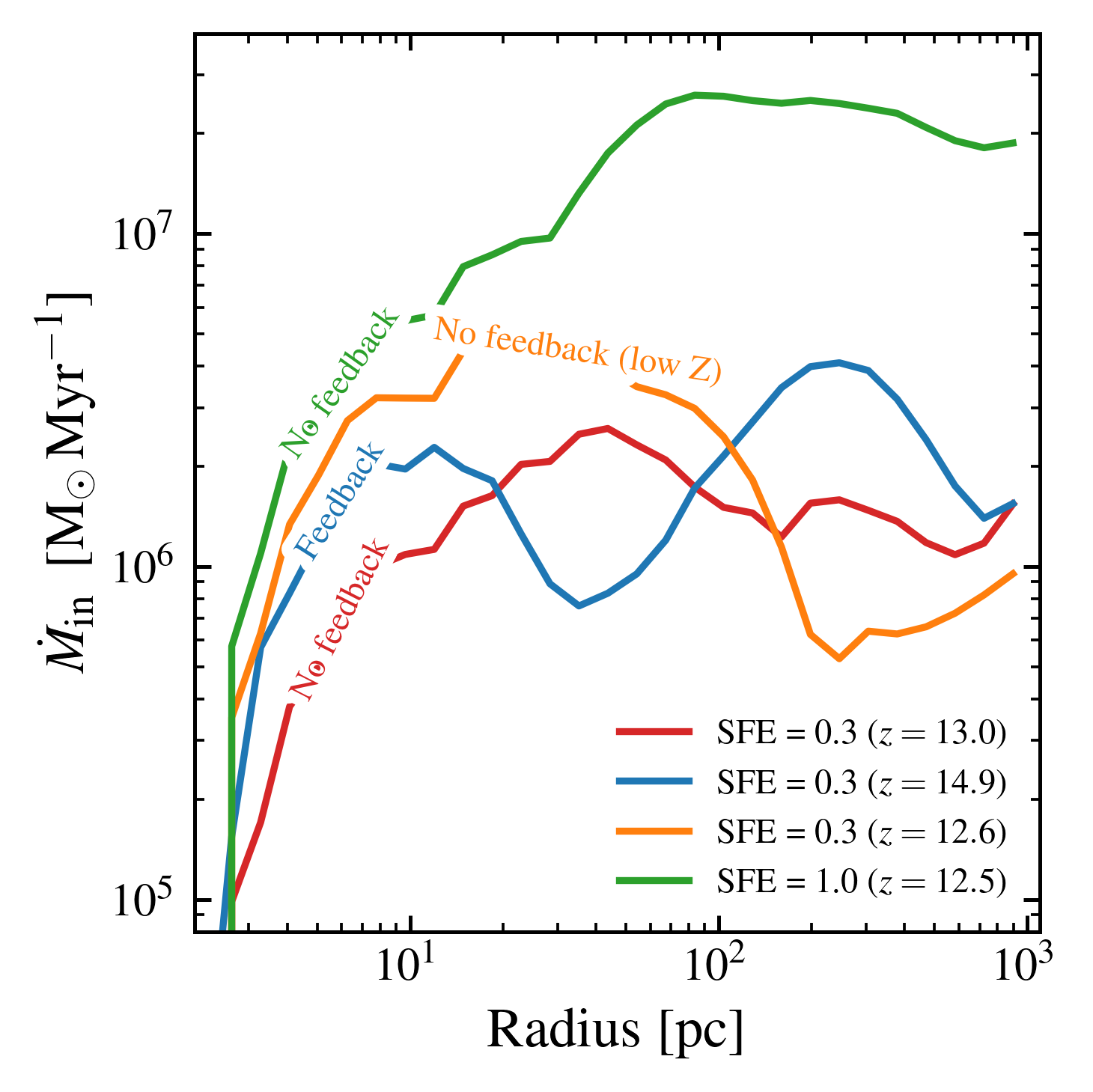}
    \caption{Radial profiles of the gas mass inflow rate ($\dot{M}_{\rm in}$). The solid curves illustrate the accretion behavior across the four simulation models: SFE $30\%$ (red), SFE $30\%$ with feedback (blue), SFE $30\%$ with low metallicity (orange), and SFE $100\%$ (green).}
    \label{fig:gasInflowProfiles}
\end{figure}

\subsection{Mass transport and formation of a central massive object}\label{timescale}

In the following, we examine the expected transport of gas and stellar mass in these systems and its implications for the formation of a central massive object. As a starting point, the gas mass inflow rates for the different simulations are shown in Fig.~\ref{fig:gasInflowProfiles} as a function of the radius. On scales of $10$~pc, all simulations show very large inflow rates of $1-3\times10^6$~M$_\odot$~Myr$^{-1}$. On larger scales, this inflow rate is approximately similar with some oscillations for most of the runs, and even increases further towards $2\times10^7$~M$_\odot$~Myr$^{-1}$ in the run with $100\%$ SFE. Thus, considering the gas dynamics alone, one would expect to obtain significant gas mass of $\sim10^7$~M$_\odot$ within the central $10$~pc in only $10$~Myr.

\begin{figure}[h!]
    \centering
    \includegraphics{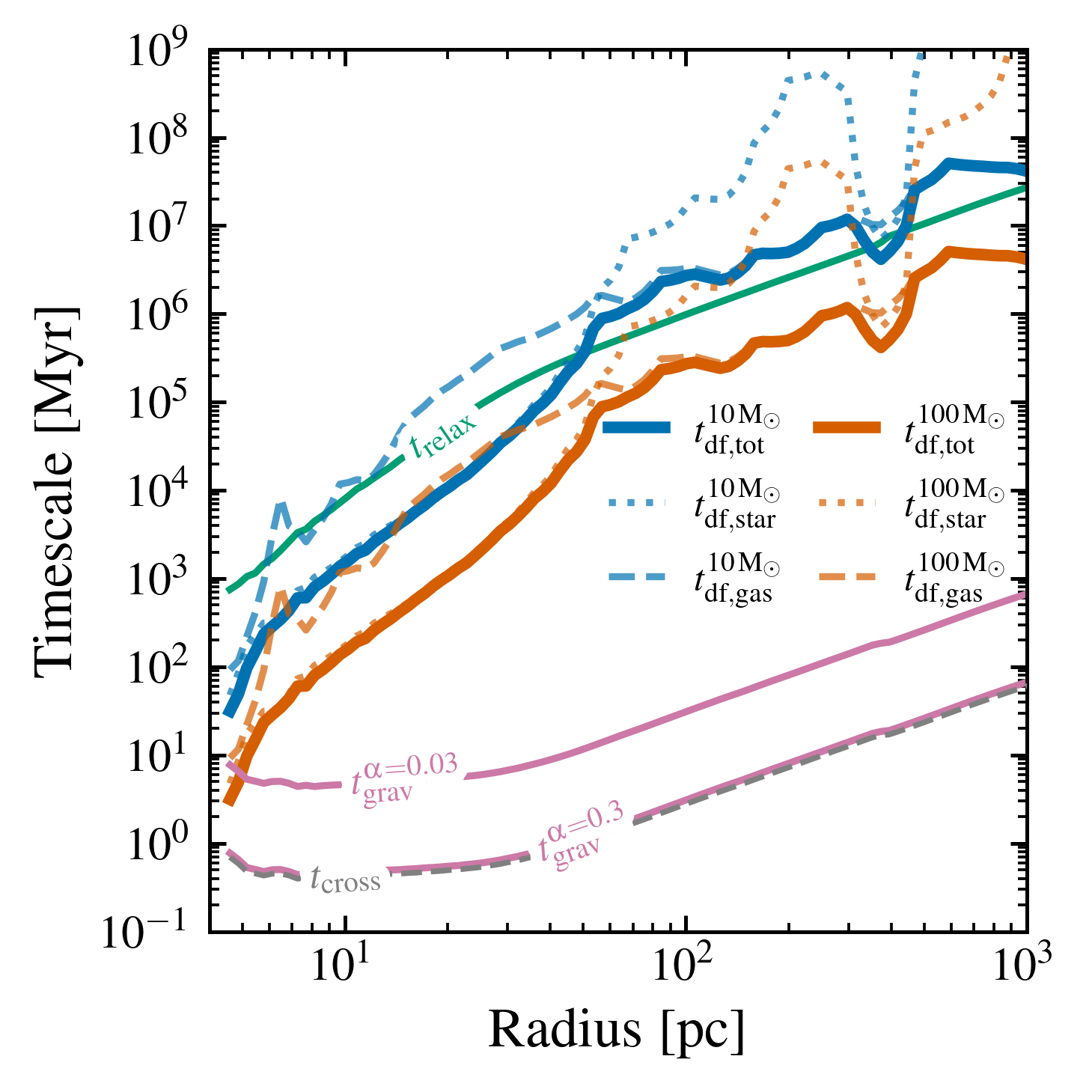}
    \caption{
    Characteristic timescales as a function of galactic radius for the simulation with $30\%$ SFE. The plot illustrates the component dynamical friction timescales exerted by the gas ($t_{\rm df, gas}$, dashed lines, equation \ref{eq:dfGas}) and the stars ($t_{\rm df, star}$, dotted lines, equation \ref{eq:dfStars}) on migrating objects of $10~{\rm M}_{\odot}$ (blue) and $100~{\rm M}_{\odot}$ (orange). The total combined dynamical friction timescale ($t_{\rm df, tot}$, equation \ref{eq:dfTotal}) is indicated by the thick solid lines. Reference timescales include the stellar crossing time ($t_{\rm cross}$, solid gray, equation \ref{eq:crossTime}), the two-body relaxation time ($t_{\rm relax}$, solid green, equation \ref{eq:relaxTime}), and the gravitational torque mass transport times ($t_{\rm grav}$, equation \ref{eq:gravTorquesTime}) for $\alpha=0.03$ and $\alpha=0.3$.)}
    \label{fig:timescales}
\end{figure}

Subsequently, we consider the transport of the stellar mass to the central region. The timescale for a massive object of mass $m_{*}$ to sink toward the center due to interactions with background stars is given by 
\begin{equation}
t_{\rm df, stars} =  \frac{\sigma^3 }{4\pi G^2m_{\star}\, \rho_{\rm stellar} \ln(0.4N)},\label{eq:dfStars}
\end{equation}
 where $\rho_\star$ is the local stellar mass density,   $\sigma = \sqrt{\frac{G(M_{\rm stellar}+M_{\rm gas})}{R}}$ the local velocity dispersion derived from the virial equilibrium, $N$ the number of stars (assuming a \citet{Kroupa2001} initial mass function) in radius $R$, and $G$ the gravitational constant. Similarly, the time scale for the drag exerted by the gaseous medium on the star with mass $m_\star$  is 
 \begin{equation}
t_{\rm df, gas} = \frac{\sigma^3}{4\pi G^2 m_{\star} \rho_{\rm gas} \ln{(0.4N)}},\label{eq:dfGas}
 \end{equation} 
 where $\rho_{\rm gas}$ is the local gas mass density, and again, $\sigma = \sqrt{\frac{G(M_{\rm stellar}+M_{\rm gas})}{R}}$. The total dynamical friction timescale ($t_{\rm df, tot}$) is then given by the inverse of the sum of the contributions from the stellar and gaseous components, that is 
 \begin{equation}
 t_{\rm df, tot}^{-1}= t_{\rm df, stars}^{-1}+t_{\rm df, gas}^{-1}.\label{eq:dfTotal}
 \end{equation}
We provide these timescales for the simulation with $30\%$ SFE in Fig.~\ref{fig:timescales} as a function of radius, considering stars of $10$~M$_\odot$ and $100$~M$_\odot$. We find that dynamical friction is dominated by the stellar component rather than gas in the inner $\sim 50$ pc.  Consequently, the total dynamical friction timescales are driven by the stellar background, yielding timescales of order $\sim10^3$~Myr at $10$~pc for $10$~M$_\odot$ stars, and even shorter timescales of $\sim10^2$~Myr for $100$~M$_\odot$. On those timescales, the massive stars are expected to become stellar mass BHs, but their accretion towards the center may likely still contribute towards the feeding and growth of a central massive object or towards the build-up of a dark core around such an object. The dynamical friction time for  $10$~M$_\odot$ objects remains less than $1$~Gyr out to radii of $\sim10$~pc, and even out to radii of $\sim30$~pc for $100$~M$_\odot$ stars, indicating a potentially large mass reservoir of $\sim10^6$~M$_\odot$ from which such stellar mass BHs could be accreted.

For comparison with other dynamical timescales, we introduce the crossing time ($t_{\rm cross}$), defined as the time required for a star to traverse the characteristic radius (R), given as
\begin{equation}
    t_{\rm cross} = \frac{R}{\sigma}.\label{eq:crossTime}
\end{equation}
As we can see in Fig.~\ref{fig:timescales}, the crossing time is of the order $0.4-0.5$~Myr up to radii of about $30$~pc and then gradually increases up to $\sim60$~Myr at $1$~kpc. The relaxation timescale $(t_{\rm relax})$ is given as
\begin{equation}
    t_{\rm relax} = 0.1 \frac{N}{\ln{\gamma N}}t_{\rm cross},\label{eq:relaxTime}
\end{equation}
 where we adopted $\gamma=0.4$, and the number of stars $N$ given by $N=M_{\rm stellar}/\langle m_*\rangle$, with the average mass $\langle m_*\rangle=0.38$\,M$_\odot$  assuming  a Kroupa IMF  with $m_{\rm min}=0.01$\,M$_\odot$ and $m_{\rm max}=150$\,M$_{\odot}$. Adopting here the Kroupa IMF likely translates into an upper limit for the relaxation time, as it is  conceivable for the IMF to be top-heavy at high redshift and lower metallicity. We thus find typical relaxation times of $\sim4\times 10^3$~Myr even at $10$~pc, which increase significantly towards larger radii. The latter implies that evolution may happen due to mass segregation or due to the interaction between gas and stars, but not as a result of two-body relaxation. 
 
\begin{figure}
    \centering
    \includegraphics[scale=0.40]{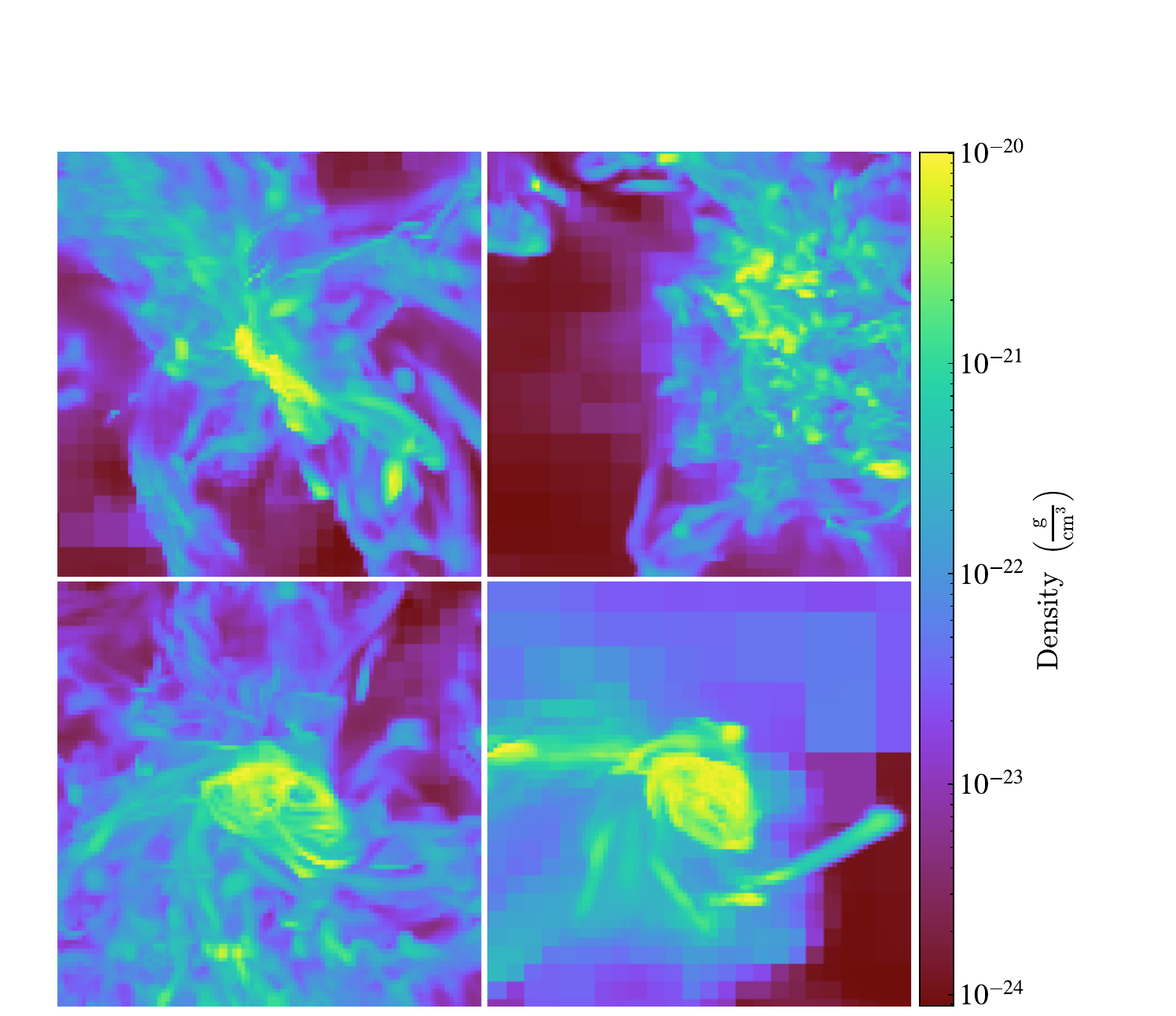}
    \caption{Density projections showing the average gas density along the line of sight for all four runs for the central 500 pc region. Top left: SFE of $30\%$; top right: SFE of $30\%$ and feedback; bottom left: SFE of $30\%$ and Z$=0.01$\,Z$_\odot$ and bottom right: SFE of $100\%$ with no feedback. }
    \label{fig:GasDensityProjection}
\end{figure}

We estimate the mass inflow rate due to gravitational torques driven by the self-gravity of the system as \citep{Lynden1972, Gammie2001, Escala2006, Escala2007, Hopkins2011, Inayoshi2014,LatifSchleicher2015} 
\begin{equation}
\dot{M}_{G}\sim3\alpha_{\rm grav}\frac{v_{\rm rot}^3}{G},\label{eq:gravitationalTorquesRates}
\end{equation}
where the rotational velocity as a function of radius has been extracted from the simulations and is provided in Fig.~\ref{fig:rotvel} in the appendix. These torques arise from gravitational instabilities within the central disk-like structures that naturally form in our simulations and are visible in the gas density projections in Figure \ref{fig:GasDensityProjection}. To capture this mechanism, we consider the $\alpha$-viscosity parameter to be in the range $\alpha_{\rm grav}\sim 0.03-0.3$, which is typical for gravitational instabilities \citep[e.g.,][]{Escala2006, Inayoshi2014, LatifSchleicher2015}.

The relevant timescale for mass transport to the center is then given as
\begin{equation}
    t_{\rm grav} =  \frac{M_{\rm gas}}{\dot{M}_G}.\label{eq:gravTorquesTime}
\end{equation}

This timescale presents a lower limit for the migration of bodies with mass $M_b$ to the center as a result of gravitational torques. In case the mass of the body under consideration considerably exceeds the mass of the disk that gravitationally interacts with that body, the timescale could also be larger. However, as here we only consider stellar-mass objects, the latter should not be a problem and we will thus more generically adopt the timescale from Eq.~\ref{eq:gravTorquesTime}. In the presence of magnetic fields, additional torques could further aid in driving mass into the center \citep{Latif2016d, Begelman2023,Latif2023, Latif2023b, Begelman2023, Diaz2024}, although we will not analyze that here in detail.

We now assess how the above mentioned timescales contribute to the mass transport into the center. The corresponding mass transport rate for $10$~M$_\odot$ stars due to dynamical friction is then
\begin{equation}
\dot{M}_{10}(r)=\epsilon_{10}\frac{ M_*(r)}{t_{\rm df,10}}, \label{eq:massTransport10}
\end{equation}
where $\epsilon_{10}\approx0.19$ is the mass fraction of stars with at least $10$~M$_\odot$ (for the assumed Kroupa IMF) and ${t_{\rm df,10}}$  the dynamical friction time for $10$~M$_\odot$ stars. By analogy, we define\begin{equation}
\dot{M}_{100}(r)=\epsilon_{100}\frac{ M_*(r)}{t_{\rm df,100}}, \label{eq:massTransport100}
\end{equation}
with $\epsilon_{100}\approx 0.019$  the mass fraction of stars with at least $100$~M$_\odot$ and ${t_{\rm df,100}}$ the dynamical friction time for $100$~M$_\odot$ stars.

For gravitational torques, on the other hand, the timescale does not depend on the mass of the objects themselves, as long as that mass is smaller than the disk mass in the local environment. We therefore have\begin{equation}
\dot{M}_{\rm grav}(r)=\epsilon_{10}\frac{ M_*(r)}{t_{\rm grav}}, \label{eq:dotMgrav}
\end{equation}
where we consider stars with at least $10$~M$_\odot$ as potentially relevant contributions. 

For the simulation with $30\%$ SFE, the timescales are provided in Fig.~\ref{fig:timescales} as a function of radius. We find that gravitational torques are particularly efficient out to radii of $\sim30$~pc with expected timescales of $0.9-10$~Myr depending on the adopted $\alpha$ parameter, and even on larger scales only slightly increasing up to a value of $\sim10^3$~Myr at $1$~kpc. Nonetheless, as we saw in Fig.~\ref{fig:MassProfiles}, this still corresponds to a significant reservoir of stellar mass.

\begin{figure}[h!]
    \centering
    \includegraphics{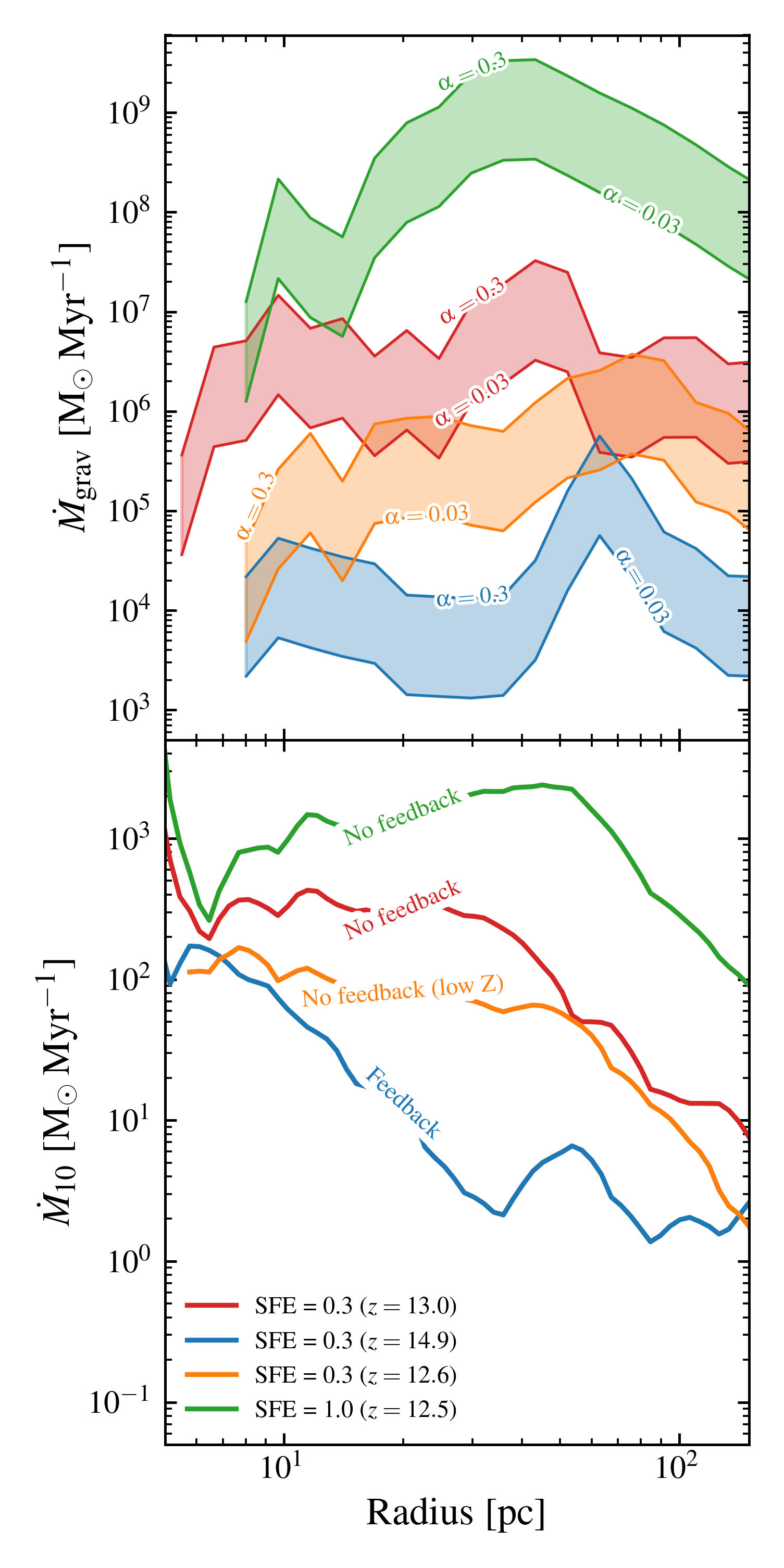}
    \caption{Radial profiles of the transport rates (equation \ref{eq:dotMgrav}, top panel)  and stellar mass transport rates (equation \ref{eq:massTransport10}, bottom panel). The curves compare the transport efficiencies across the varied simulation setups: the SFE $30\%$,  no feedback and $0.1$\,Z/Z$_\odot$ (red), the SFE $30\%$ model with feedback and $0.1$\,Z/Z$_\odot$ (blue), the SFE  $30\%$ model with no feedback and $0.01$\,Z/Z$_\odot$(orange), and the SFE  $100\%$ no feedback and $0.1$\,Z/Z$_\odot$  model (green).} \label{fig:massTransportRate}
\end{figure}

In Fig.~\ref{fig:massTransportRate}, we compare the mass transport rates due to gravitational torques and dynamical friction for the different simulations. The mass transport rate from gravitational torques varies more strongly among the different simulations, both due to the difference in stellar mass between the simulations (Fig.~\ref{fig:MassProfiles}) and the strong dependence on the rotational velocity (Fig.~\ref{fig:rotvel}), from minimum values of $\sim10^4$~M$_\odot$~Myr$^{-1}$ (simulation with supernova feedback) up to peak values in the range of $10^8-10^9$~M$_\odot$~Myr$^{-1}$ (simulation with $100\%$ SFE). Typical intermediate values are in the range of $10^5-10^6$~M$_\odot$~Myr$^{-1}$, and the dependence on the distance from the center is not very strong, even if some peaks and oscillations may occur. Within $10$~Myr, we could conservatively expect a mass accumulation of at least $\sim10^5$~M$_\odot$, though it could also be considerably higher and may depend both on the physics and the environment.

For dynamical friction, we find typical mass transport rates of $10^2-10^3$~M$_\odot$~Myr$^{-1}$ on the $10$~pc scale, rather similar for the 4 different runs, though they may vary by a few orders of magnitude on somewhat larger scales. Within $10$~Myr, we would expect a mass of $10^3-10^4$~M$_\odot$ to accumulate in the center. A summary of the expected mass transport rates due to different processes within the different simulations on a $10$~pc scale is given in Table~\ref{tab:transport_rates}.

\begin{deluxetable*}{lccccc}
\tabletypesize{\footnotesize}
\tablecaption{Mass transport rates at $10\,{\rm pc}$. \label{tab:transport_rates}}
\tablehead{
  \colhead{Simulation} & \colhead{redshift} &
  \colhead{$\dot{M}_{\rm in}$ } &
  \colhead{$\dot{M}_{10}$} &
  \colhead{$\dot{M}_{100}$} &
  \colhead{$\dot{M}_{\rm grav}$} \\
  \colhead{} & \colhead{} &
  \colhead{[M$_\odot\,\rm Myr^{-1}$]} &
  \colhead{[M$_\odot\,\rm Myr^{-1}$]} &
  \colhead{[M$_\odot\,\rm Myr^{-1}$]} &
  \colhead{[M$_\odot\,\rm Myr^{-1}$]}
}
\startdata
SFE = 30\% & $13.0$ & $1.09 \times 10^{6}$ & $1.31 \times 10^{2}$ & $1.35 \times 10^{2}$ & $1.37 \times 10^{6}$ -- $1.37 \times 10^{7}$ \\
SFE = 30\% LowZ & $14.9$ & $3.21 \times 10^{6}$ & $4.13 \times 10^{1}$ & $4.26 \times 10^{1}$ & $3.15 \times 10^{4}$ -- $3.15 \times 10^{5}$ \\
SFE = 30\% FB & $12.6$ & $2.01 \times 10^{6}$ & $2.71 \times 10^{1}$ & $2.80 \times 10^{1}$ & $3.45 \times 10^{3}$ -- $3.45 \times 10^{4}$ \\
SFE = 100\% & $12.5$ & $5.51 \times 10^{6}$ & $3.46 \times 10^{2}$ & $3.57 \times 10^{2}$ & $1.95 \times 10^{7}$ -- $1.95 \times 10^{8}$ \\
\enddata
\tablecomments{Mass transport rates evaluated at a characteristic radius of $10\,{\rm pc}$. Gas inflow rates ($\dot{M}_{\rm in}$) represent radial fluxes extracted from the simulations. Dynamical friction rates ($\dot{M}_{10}$ and $\dot{M}_{100}$, equation \ref{eq:massTransport10}) are evaluated for objects of $10\,{\rm M}_\odot$ and $100\,{\rm M}_\odot$, respectively. Gravitational torque mass transport rates ($\dot{M}_{\rm grav}$, equation \ref{eq:gravitationalTorquesRates}) are presented as a range corresponding to $\alpha = 0.03 \text{--} 0.3$.}
\end{deluxetable*}

In summary, during a timescale of $10$~Myrs, we could expect a mass of $10^7$~M$_\odot$ to reach the center of the galaxy due to gas inflow, an additional mass corresponding to massive stars and stellar mass BHs in the range of $10^5-10^9$~M$_\odot$ due to gravitational torques, as well as an additional mass of $10^3-10^4$~M$_\odot$ as a result of dynamical friction. For a conservative estimate, we here take the mass expected from the gas inflow and consider an efficiency factor of $10\%$ due to the possible presence of feedback during the formation of a massive BH. This would result in an expected BH mass of $10^6$~M$_\odot$, consistent with typical BH mass estimates for LRDs based on the observed infrared luminosities \citep[e.g.][]{Liempi2026}. In case of additional efficient feeding through gravitational torques, the mass could be higher by a relevant amount. It is also worth noting that even just dynamical friction itself, while neglecting all other processes, could lead to the formation of a massive object or at least a massive cluster of BHs of $\sim10^4$~M$_\odot$. While of course the interaction of the different components that are being transported to the center will require further investigation, the significant mass transport via 3 independent channels nonetheless provides a strong case for the formation of a massive BH and the very likely formation of an AGN.

\section{Discussion and conclusions} \label{sec:summary}

In this paper, we have presented zoom-in cosmological hydrodynamics simulations exploring the regime of high SFEs and confined feedback, as motivated by JWST observations \citep{Chworowsky2024, Somerville2025}, cloud-scale simulations \citep{Kim2018, Lancaster2021, Menon2025} and theoretical considerations \citep{Haid2018, Dekel2023, Dekel2025, Yajima2025}. Specifically, we consider simulations with SFEs of $30\%$ and $100\%$. As a typical metallicity in this regime, we adopt $0.1$~Z$_\odot$, but we also consider a lower-metallicity case with $0.01$~Z$_\odot$. Most of the runs are without feedback, while in one simulation supernova feedback is explicitly included. We initialize our simulations at $z=200$ with cosmological initial conditions and consider a simulation box of $37.3$~Mpc, employing a top grid resolution of $512^3$, an additional refinement level in the Lagragian volume of the most massive halo as well as $10$ dynamical levels of refinement, resolving scales down to $\sim4$~pc. 

In the most massive halo with a DM mass of $\sim3\times10^{10}$~M$_\odot$ at $z=11$, massive and compact stellar systems form in all simulations, with stellar masses of $\sim10^7-4\times10^8$~M$_\odot$ within $\sim200$~pc, and gas masses of $8\times10^6-3\times10^7$~M$_\odot$. 
These systems thus represent ideal progenitors for the lower-mass limit of the LRD population when adopting the interpretation of very compact stellar systems \citep{GREENE2024, MATTHEE2023}, and also fulfill the compactness criteria implied by the observations. We emphasize here that these systems form naturally form cosmological initial conditions under the assumption of high SFE.
In the simulation with supernova feedback, the build-up of stellar mass has been delayed, but it still shows an enclosed stellar mass of $10^7$~M$_\odot$ in $200$~pc at $z=14.9$, and is thus likely to reach larger stellar masses during their further evolution. While our primary simulations explore the limit of confined supernova feedback, it is worth noting the potential role of early stellar feedback mechanisms, such as stellar winds, photoionization, and radiation pressure from UV and IR scattering. In typical galactic environments, these processes can drive local outflows and self-regulate the local SFE. However, high core densities reaching $10^4$ to $10^8$\,M$_\odot$\,pc$^{-3}$, the associated large optical depths and deep potential wells are expected to severely confine these early feedback mechanisms. Consequently, their global impact on preventing mass accumulation is likely to be small, consistent with the confinement of feedback seen in dense cloud-scale simulations  \citep[e.g.,][]{Haid2018}.

We in addition find in all of these simulations very efficient mass transport into the central region, with typical gas inflow rates of $1-5\times10^6$~M$_\odot$~Myr$^{-1}$, transport of massive stars and stellar mass BHs into the center via gravitational torques with rates ranging from $5\times10^3$~M$_\odot$~Myr$^{-1}$ up to $2\times10^7$~M$_\odot$~Myr$^{-1}$ as well as mass transport rates due to gas and stellar dynamical friction of $\sim3\times10^{3}-3\times10^{4}$~M$_\odot$~Myr$^{-1}$. Over a timescale of $10$~Myr, it is thus highly plausible that at least $10^7$~M$_\odot$ will arrive in the central region, even if we just consider the gas inflow. The interaction of gas with stellar dynamics has been at least partially looked at and is likely to favor the formation of a very massive object \citep{Boekholt2018,Tagawa2020, Chon2020, Chon2025,Kroupa2020,   Das2021a, Das2021b,  Schleicher2022, Reinoso2023, Gaete2024, Solar2025}. We performed cosmological simulations to model the assembly and evolution of compact galaxies hosting multiple star clusters in realistic cosmological environments. In contrast, direct N-body approaches, such as \cite{Rantala2024}, focus on the internal cluster dynamics and stellar runaway mergers of individual systems. To bridge this gap, future work will map the outputs of cosmological simulations onto direct N-body codes, incorporating gas dynamics within the AMUSE framework.

However, it is necessary to consider the limitations of standard dynamical friction estimates in the highly clumpy environments characteristic of high-redshift galaxies. Our simulated systems exhibit pronounced substructure and clumpiness. While one might intuitively expect such clumpiness to disrupt or reduce the efficiency of orbital decay, recent studies have shown that clumpy environments can actually drive more efficient central mass accumulation and nuclear fueling \citep[e.g.,][]{DeGraf2017}. To verify the efficiency of cluster sinking in our regime, we analyzed the properties of massive clumps (up to $5 \times 10^7$\,M$_\odot$) across our different models. These clumps are at relatively short distances ($\sim 100$ pc) and have sinking timescales typically ranging between 2 and 40 Myr, even when accounting for the delays introduced by stellar feedback. This rapid orbital decay underscores that these massive clusters will efficiently sink to the center, providing the necessary high-density conditions for rapid nuclear growth and potential black hole seeding. Nonetheless it is clear that we are considering here a both physically and computationally challenging regime and more detailed investigations about the outcome of such an inflow of gas, stars and stellar mass BHs will be required. We adopt here a conservative efficiency estimate of $10\%$ of the BH from the inflowing material, suggesting that a massive BH of at least $10^6$~M$_\odot$ would be expected to form. Such a BH mass is well within the typically expected BH masses when considering the infrared luminosities of LRDs \citep{Liempi2026}. The resulting BH would be surrounded by gas, stars and even stellar mass BHs, naturally leading to the formation of an AGN with a high probability to be accompanied by tidal disruption events (TDEs) as well as extreme mass ratio inspirals (EMRIs). It is thus at least very likely that the resulting object will provide important ingredients that are expected to be relevant in the context of LRDs. 

Our simulations here also indicate a possible pathway to unify the dense stellar system interpretation with the AGN interpretation of LRDs. While we find the formation of such dense systems to happen very naturally, the processes within these systems also naturally lead to the formation of an AGN on rather short timescales. Having a dense stellar system or an AGN thus should not be considered as mutually exclusive alternatives, but rather the dense and compact system provides a natural pathway that may even be required to explain the formation and presence of a central massive black hole. 

Some similar results have been found as part of the VELA simulations following the larger-scale evolution with a maximum resolved scale of $25$~pc, finding the formation of wet-compaction events in DM halos of $\sim10^{10}$~M$_\odot$ \citep{Ceverino2014, Zolotov2015, Lapiner2023}. Such wet compactions provide similarly compact systems and it is conceivable that they would show very similar dynamics as the systems modeled here in case of a comparable resolution.   

In future studies, both the actual formation process of a massive BH from this configuration should be investigated in more detail,  the expected appearance and spectral features of the system and its subsequent long-term evolution. So far the system that formed here is at least compatible with several of the suggested pathways to form LRDs. Independent of the question on whether these systems evolve into LRDs, we also highlight that they present a very efficient pathway to form quite massive BHs and provide an interesting astrophysical laboratory including processes such as TDEs and EMRIs with potentially significant source counts. The formation and evolution of such configurations thus clearly requires more investigation in the future.

\begin{acknowledgments}

ML gratefully acknowledges support from ANID/DOCTORADO BECAS CHILE 72240058. MAL thanks the UAEU for funding via UPAR grant No. G00005454. DRGS thanks for funding via the ANID BASAL project FB21003 and the Alexander von Humboldt Foundation.
\end{acknowledgments}

\begin{contribution}
All authors contributed equally to this  work.
\end{contribution}

\software{astropy \citep{2013A&A...558A..33A,2018AJ....156..123A,2022ApJ...935..167A},  
          ENZO \citep{enzo}
          }

\appendix

\section{Rotational velocity profiles and metallicity}

In this appendix, we provide additional details regarding the kinematic and chemical properties of the gas in our simulated systems, specifically focusing on the rotational velocity profiles across all runs and the metallicity distribution in the presence of stellar feedback.

\begin{figure}[!h]
    \centering
    \includegraphics{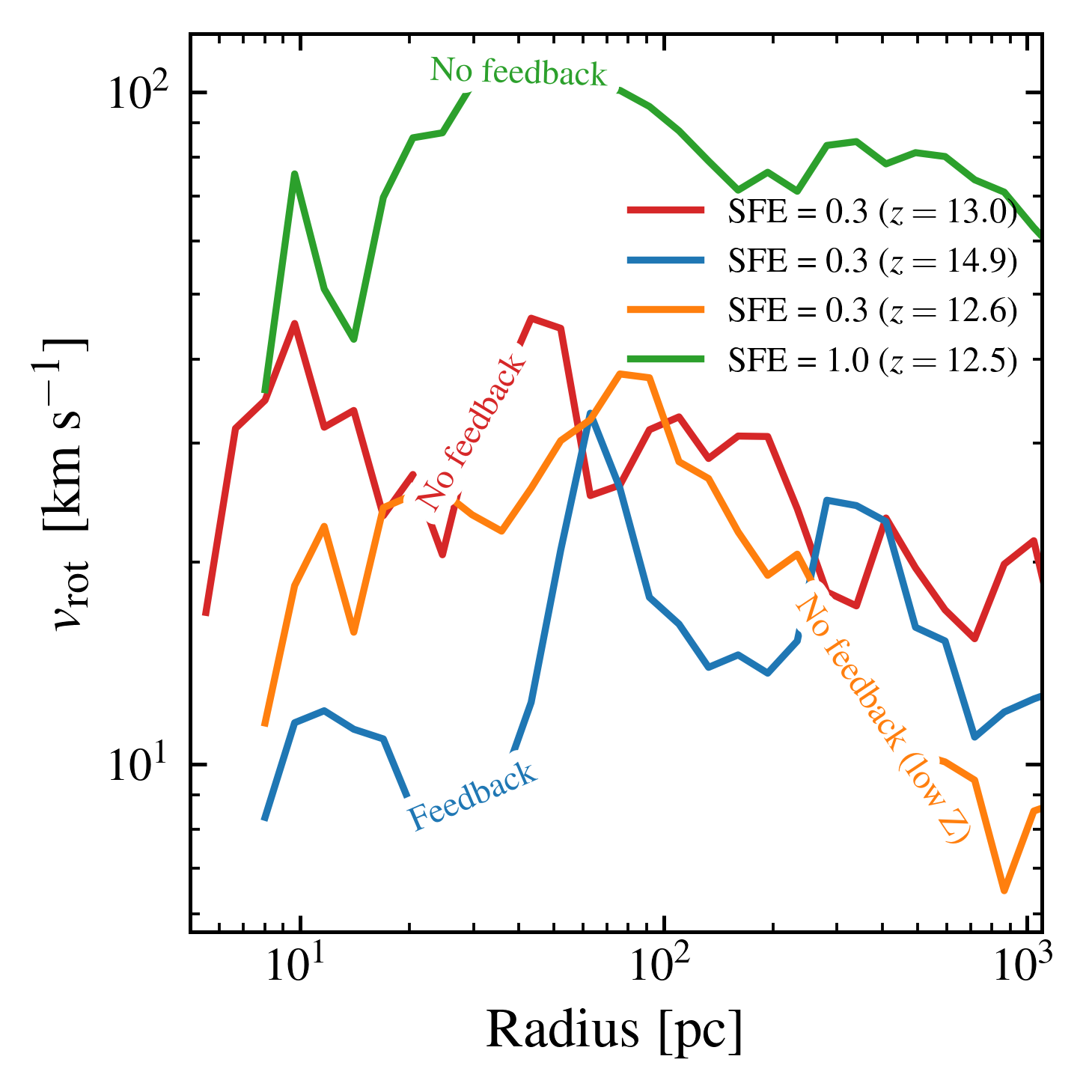}
    \caption{Rotational velocity ($v_{\rm rot}$) as a function of radius for the different ENZO simulations. Higher SFE leads to greater rotational velocities due to a deeper central potential, while stellar feedback severely disrupts coherent rotation in the inner $50\,{\rm pc}$.}
    \label{fig:rotvel}
\end{figure}

Figure~\ref{fig:rotvel} shows the rotational velocity ($v_{\rm rot}$) of the gas as a function of radius. The kinematic signatures are highly dependent on both the star formation efficiency (SFE) and the inclusion of feedback. The extreme SFE $= 1.0$ run without feedback (green line) exhibits the highest rotational velocities, peaking near $100\,{\rm km\,s^{-1}}$ between $60$ and $100\,{\rm pc}$. This reflects the deep central potential well formed by the rapid, unimpeded accumulation of gas and stars. Reducing the SFE to $30\%$ (red and orange lines) correspondingly lowers the rotational support, with velocities generally fluctuating between $10$ and $50\,{\rm km\,s^{-1}}$. 

The inclusion of stellar feedback (blue line) drastically alters the inner kinematic structure of the galaxy. Within the central $50\,{\rm pc}$, $v_{\rm rot}$ drops significantly below $10\,{\rm km\,s^{-1}}$, demonstrating that energy and momentum injected by supernovae effectively disrupt the coherent rotational disk, injecting turbulence and driving outflows that prevent the gas from settling into a rotationally supported state at small radii.

\begin{figure}[!h]
    \centering
    \includegraphics[width=0.8\columnwidth]{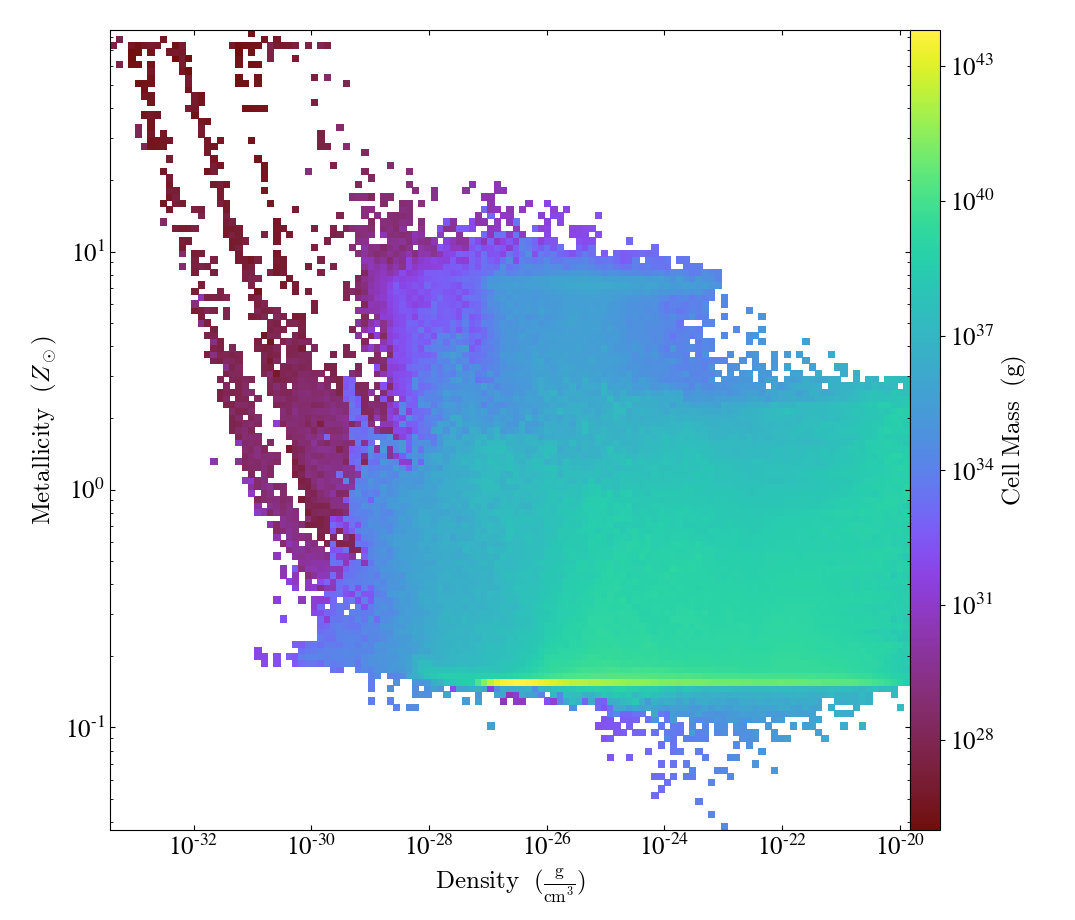}
    \caption{Gas metallicity ($Z_\odot$) as a function of gas density ($\rm g\,cm^{-3}$) for the feedback run with SFE $= 30\%$. The color bar denotes the total gas cell mass. The presence of highly metal-enriched gas at very low densities indicates the presence of feedback-driven outflows.}
    \label{fig:metallicity}
\end{figure}

To further illustrate the impact of this feedback, Figure~\ref{fig:metallicity} presents a 2D phase diagram of gas metallicity versus density for the SFE $= 30\%$ feedback run. The color mapping indicates the total cell mass in each phase space bin. While the bulk of the dense gas mass (yellow/green horizontal band extending to $10^{-20}\,{\rm g\,cm^{-3}}$) remains at a relatively uniform, sub-solar metallicity ($\sim 0.15\,{\rm Z}_\odot$), there is a distinct vertical scatter of highly enriched gas. 

Supernova feedback heavily pollutes the surrounding medium, driving gas up to and beyond solar metallicities ($> 10\,{\rm Z}_\odot$). This highly enriched gas is primarily found at lower densities ($< 10^{-28}\,{\rm g\,cm^{-3}}$).

\bibliography{ref}{}
\bibliographystyle{aasjournalv7}

\end{document}